\documentclass[proof]{WileyASNA-v1}

\articletype{Article Type}%

\received{18 October 2019}
\revised{XXXX}
\accepted{XXXX}
\usepackage{graphicx}

\begin{document}

\title{The Athena space X-ray Observatory and the astrophysics of hot plasma \protect\thanks{Paper presented on behalf of the Athena Science Study Team composed of D. Barret, A. Decourchelle, A.C. Fabian, M. Guainazzi, J.W. den Herder, H. Matsumoto, K. Nandra, L. Piro, R. Smith, R. Willingale}}

\author[1]{Didier Barret*}

\author[2]{Anne Decourchelle}

\author[3]{Andy Fabian}

\author[4]{Matteo Guainazzi}

\author[5]{Kirpal Nandra}

\author[6]{Randall Smith}
\author[7]{Jan-Willem den Herder}
\authormark{Barret \textsc{et al}}

\address[1]{\orgname{IRAP}, \orgaddress{\state{Université de Toulouse; CNRS; Institut de Recherche en Astrophysique et Plan\'etologie; 9 Avenue du colonel Roche, BP 44346, F-31028 Toulouse cedex 4}, \country{France}}}

\address[2]{\orgname{CEA}, \orgaddress{\state{Service d'Astrophysique, CEA Saclay, F-91191 Gif-sur-Yvette Cedex}, \country{France}}}

\address[3]{\orgname{IOA}, \orgaddress{\state{Institute of Astronomy, University of Cambridge, Madingley Road, Cambridge, CB3 0HA}, \country{United Kingdom}}}

\address[4]{\orgname{ESA}, \orgaddress{\state{ESTEC, Keplerlaan 1, 2201AZ, Noordwijk}, \country{The Netherlands}}}

\address[5]{\orgname{MPE}, \orgaddress{\state{Max-Planck-Institut fur extraterrestrische Physik, Giessenbachstrasse 1, 85748 Garching}, \country{Germany}}}

\address[6]{\orgname{Smithsonian Astrophysical Observatory, 60 garden street}, \orgaddress{\state{Cambridge, MA 02138}, \country{United States}}}
\address[7]{\orgname{SRON, Netherlands Institute for Space Research, Sorbonnelaan 2, 3584 CA Utrecht}, \orgaddress{\country{The Netherlands}}}

\corres{*Corresponding author: Didier Barret. \email{dbarret@irap.omp.eu}}

%\presentaddress{This is sample for present address text this is sample for present address text}

\abstract{The properties (temperature, density, chemical composition, velocity) of hot astrophysical plasma and the physical processes affecting them (heating/cooling, turbulence, shocks, acceleration) can be probed by high resolution X-ray spectroscopy, to be complemented by high spatial resolution imaging. The paper presents a status of the ESA's Advanced Telescope for High Energy Astrophysics (Athena) mission, particularly focusing on the science performance of its two focal plane instruments for the studies of extended X-ray sources: the Wide Field Imager (WFI) and the X-ray Integral Field Unit (X-IFU). This paper then provides a brief summary of the breakthroughs expected with \textit{Athena} on the astrophysics of hot plasma, building on the vast heritage of the discoveries and revolutionary results obtained by {\it Chandra} and XMM-Newton in this field. As of November 12th, 2019, Athena successfully concluded its feasibility study, and has since then moved into the definition phase, with a launch date scheduled in the early 2030s.}


\fundingInfo{DB acknowledges financial support from the French Space Agency (CNES).}

\maketitle

\footnotetext{\textbf{Abbreviations:} Athena: Advanced Telescope for High Energy Astrophysics, WFI: Wide Field Imager, X-IFU: X-ray Integral Field Unit}

\section{Introduction}
Hot X-ray emitting astrophysical plasma arise in various settings in the Universe: from hot gas trapped in dark matter potentials, to galactic halos and winds, down to supernova remnants. XMM-Newton and \textit{Chandra} have provided a wealth of X-ray spectral data, probing the properties of these hot plasma and the processes affecting their physical properties, as highlighted during this conference. Building on this great heritage, in the next decades the X-ray Imaging and Spectroscopy Mission ( \textit{XRISM}) and later \textit{Athena} will push up the study of the Hot Universe \citep{guainazzi_2018_arxiv}, providing both spatially resolved high resolution spectroscopy and wide field imaging of extended sources. Among the core science objectives of \textit{Athena} many are linked to hot astrophysical plasma, grouped in what is called the Hot Universe \cite[and references therein]{nandra_2013_arxiv,barcons_2015_jpcs,barret_sf2a_2013}. In particular, \textit{Athena} will enable to i) measure how gravitational energy is dissipated into bulk motions and gas turbulence in a representative galaxy cluster population, ii) determine the evolution of the gas thermodynamics during hierarchical gravitational collapse as a function of cosmic epoch by measuring the structural properties (e.g. the entropy profiles) of a sample of high mass clusters, iii) map the production and circulation of metals in the deep potential wells of massive galaxy clusters across time, iv) measure and characterize cluster feedback (bubbles, ripples, heating/cooling, radio lobes), and v) find the missing baryons pervading the cosmic web \citep[and references therein]{pointecouteau_sp_2013,ettori_sp_2013,croston_sp_2013,kaastra_2013_sp}. In addition, several so-called observatory science objectives touch also on hot astrophysical plasma in extended X-ray sources; amongst others mapping the auroral and exosphere X-ray emissions of solar system bodies, studying particle acceleration in pulsar wind nebulae, characterizing the warm/hot interstellar medium in nearby galaxies, constraining SN1a and core-collapse explosion models and shock dynamics \citep[and references therein]{branduardi_sp_2013,decourchelle_sp_2013}.

 The paper first provides an overview of the \textit{Athena} mission and presents its two focal plane instruments: the Wide Field Imager (\S \ref{wfi}) and the X-ray Integral Field Unit (\S \ref{xifu}). Both instruments have just completed successfully their Preliminary Requirement Review (PRR). Then the paper provides a brief summary of the breakthroughs expected from \textit{Athena} for both extragalactic  (\S \ref{diffextragal}) and galactic science (\S \ref{diffgal}) in the field of hot X-ray plasma, referring to the available presentations given during the conference\footnotetext{Some presentations are available at \url{https://www.cosmos.esa.int/web/xmm-newton/2019-workshop}} or the papers published in this volume.

\section{The {\it Athena} mission}
\label{athena}

{\it Athena} is the second Large-class mission of the European Space Agency (ESA) in the framework of its Cosmic Vision Program. This X-ray observatory was selected in 2014 to address ``The Hot and Energetic Universe'' science theme, aiming at answering the following questions:

\begin{itemize}

\item how does baryonic matter assemble in large-scale structures? How do these structures evolve from the formation epoch at redshift, $z$, $\ge$2 to the present day?

\item how do black holes grow and shape galaxies?
  
\end{itemize}

At the same time, {\it Athena} shall be able to ensure a large discovery space in almost all areas of modern astrophysics. According to the current science management plan, about two thirds of the {\it Athena} observation time during the 4-year nominal mission should be allocated through a competitive peer-review process open to the astronomical community worldwide. 
%In this context, it is particularly noteworthy that the {\it Athena} spacecraft and ground segment are design to respond within 4~hours with a $\ge$50\% efficiency to any astronomical event in a random position of the sky, most notably Gamma Ray-Bursts.

\begin{figure}
\centerline{\includegraphics[scale=0.21]{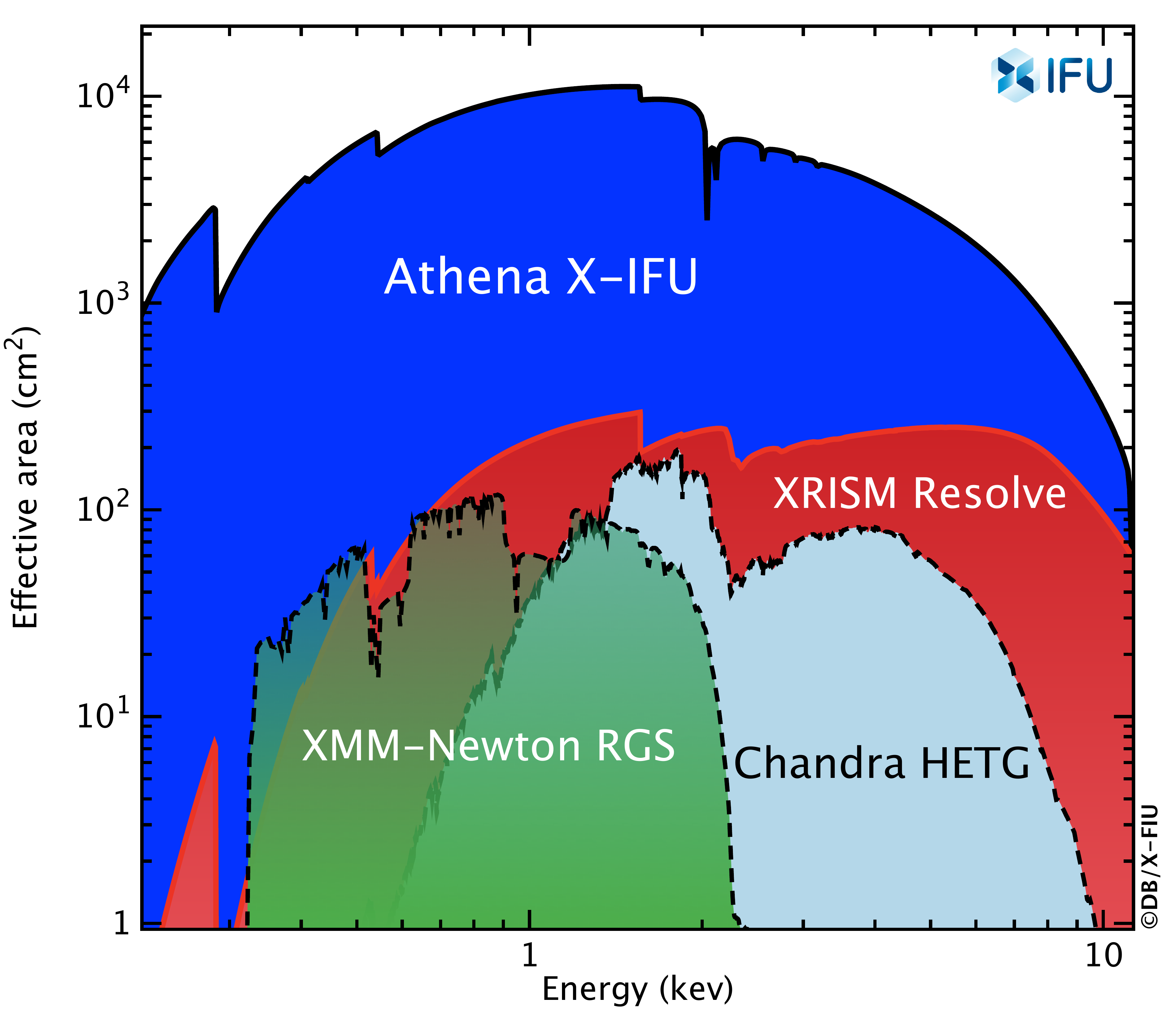} }
\caption{Effective area of \textit{Athena} X-IFU compared to the one of the  \textit{XRISM} Resolve instrument, both being imaging spectrometers. Two dispersive spectrometers currently in operation, namely the XMM-Newton/RGS and the {\it Chandra}/HETG, are also shown. Note that {\it Athena} offers imaging capability with an optics angular resolution of 5'' with a pixel size less than 5'' for X-IFU. This is to be compared with the angular resolution of 1' for the  \textit{XRISM} optics and the pixel size of 30'' for the \textit{Resolve} instrument (see Fig. \ref{perseus_xifu_resolve} to visualize what it means in terms of imaging quality).}
\label{aeff_spectro}
\end{figure}

In order to achieve its scientific goals, {\it Athena} will carry a scientific payload with an unprecedented combination of scientific performance:

\begin{itemize}

\item a single telescope based an innovative Silicon Pore Optics technology developed in Europe \citep{bavdaz2018}, with a 12 meter focal length, a 5'' Half Energy Width angular resolution, and an effective area $\ge$1.4~m$^2$ at 1~keV, and $\ge$0.25~m$^2$ at 6 keV.

\item a {\it Wide Field Imager} \citep{meidinger2018} Active Pixel Sensor, wide-field (40'$\times$40') spectral-imaging detector with an energy resolution of $\le$170~eV at 7~keV (cf. Tab.~\ref{tab:wfi} and \S~\ref{wfi})

\item an {\it X-ray Integral Field Unit} \citep{barret_2018_spie}, with an energy resolution $\le$2.5~eV up to 7 keV and spectral coverage over the whole 0.2-12~keV energy band over a 5' hexagonal field-of-view (equivalent diameter) and a $\le$5'' pixel size (cf. Tab.~\ref{tab1_xifu} and \S~\ref{xifu})
  
\end{itemize}

An ``Instrument Switching Mechanism'', based on a set of hexapods, allows each of the two instruments to be located at the mirror focus at any given time. The same mechanism allows to defocus the mirror Point Spread Function (PSF), permitting to increase the capability of the X-IFU to observe bright X-ray sources (cf. \S~\ref{xifu}). A halo orbit around L2, the second Lagrange point of the Sun-Earth, is the current baseline, with L1 still being evaluated as a back-up option contingent on a more detailed comparison of the background radiation environment.

The combination of the mirror collective area and instrument efficiency delivers a total effective area at 1~keV larger by almost two orders of magnitude than any existing or planned high-resolution X-ray spectrometers (while ensuring a significantly better energy resolution above 1~keV, see Fig.\ref{aeff_spectro}), and better by more than one order-of-magnitude than any X-ray CCD instruments in operation, or planned to be flown.

{\it Athena} has recently transitioned from Phase~A (feasibility) to Phase~B1 (preliminary definition). The successful PRR of both instruments was a critical milestone to confirm the feasibility of both instrument designs and of the definition of the interfaces with the spacecraft and the focal plane module. At system-level, the Mission Formulation Review (MFR) verified the completeness, adequacy and consistency of the mission and spacecraft preliminary design, of the interfaces, and of the Phase~B1 management and associated risk management plan; verified the completeness, adequacy and consistency of the mission and spacecraft requirement flow-down; verified the adequacy of the ESA Technology Plan to achieve the required level of technological readiness by adoption; and verified the realism of the mission schedule and cost estimate. The MFR was concluded successfully on November 12th, 2019; another major milestone in the development of Athena.

The Mission Adoption Review is current scheduled in the summer of 2021, to achieve the ultimate adoption of the mission in the ESA science program by the end of 2021. The launch is currently scheduled in the early 2030s, pending on the optimization of the schedules of both the spacecraft and the science payload. 

\section{The Wide Field Imager}
\label{wfi}
\begin{figure}[t]
\centerline{\includegraphics[clip,width=0.48\textwidth]{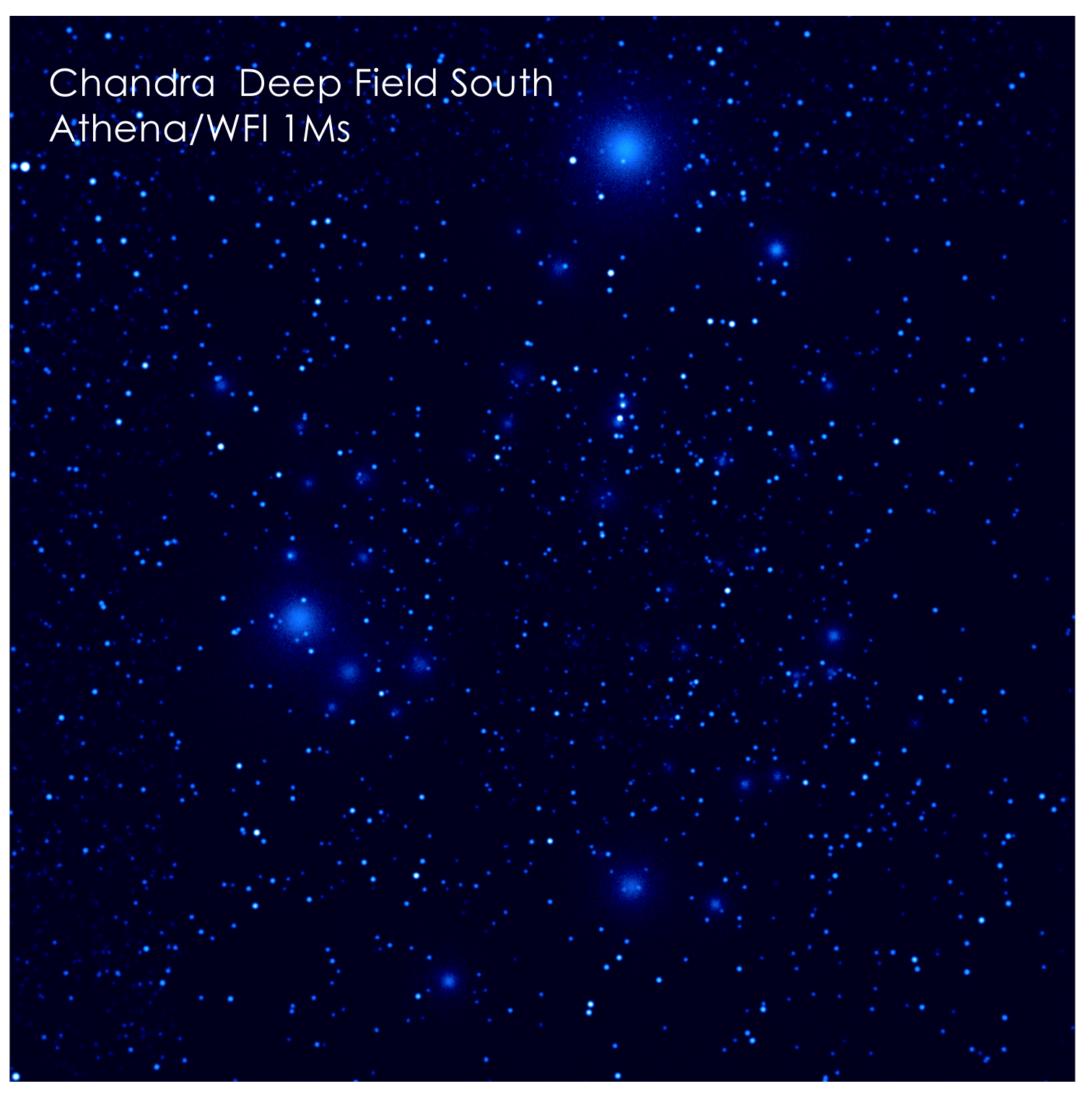}}
\caption{Deep (1Ms) survey simulation of the \textit{Chandra} Deep Field South with the WFI. This forms part of a multi-tiered survey which will address multiple science goals, for example revealing the first hot gas structures and the early supermassive black hole population. Diffuse sources present in the images are clusters of galaxies. Credit: WFI team\label{fig:wfi}}
\end{figure}

The Wide Field Imager (WFI) is a powerful spectral-imaging camera for \textit{Athena} designed to address several key elements of the Hot and Energetic Universe science theme. The instrument is based on unique Silicon DEPFET (DEPleted Field Effect Transistor) technology, which provides the WFI’s large field of view, excellent energy resolution, low noise, fast readout and high time resolution, with good radiation hardness  \citep{meidinger2018}. These capabilities, in combination with the unprecedented effective area and wide field of the \textit{Athena} telescope, will provide breakthrough capabilities in wide-field X-ray imaging spectroscopy. Observational programs with the WFI will, for example: uncover typical supermassive black hole (SMBH) activity at $z>6$, into the dark ages where the first stars and galaxies formed; perform a census of black hole activity at $z=1-4$, including for the most obscured objects, to probe the incidence, duty cycle and energetics of powerful outflows that could represent the agent of "AGN feed-back" onto the host galaxy \citep{cappi_sp_2013}. Related to extended sources, WFI will pinpoint the hot gas occupying the most massive dark matter halos at z $\sim 2$ when the first groups and clusters of galaxies formed; measure the temperature and abundances of clusters of galaxies out to their virial radius \citep{ettori_sp_2013,pointecouteau_sp_2013}. Many of these science objectives will be achieved via a multi-tiered extragalactic survey (Fig.~\ref{fig:wfi}) with sensitivity comparable to the deepest X-ray surveys currently possible, but over orders of magnitude larger areas. WFI will also perform spectral-timing measurements of extremely bright compact sources to determine the structure of their innermost accreting regions. 

\begin{center}
\begin{table}[t]%
\centering
\caption{WFI key performance requirements.\label{tab:wfi}}%
\tabcolsep=0pt%
\begin{tabular*}{20pc}{@{\extracolsep\fill}l@{\extracolsep\fill}}
\toprule
\textbf{Performance parameters}  \\
\midrule
Energy range : 0.2--15 keV \\
Spectral resolution : $<170$ eV (7 keV, end of life) \\
Field of view : $40 \times 40$ arc min\\
Pixel size : $2.2$ arc seconds \\
Non X-ray background : $5 \times 10^{-3} \rm ~counts/s/cm^2/keV$\tnote{$\dagger$}   \\
Count rate capability: 1 Crab at 95\% throughput \\
%
%col1 text & col2 text  & col3 text  & col4 text  & col5 text\tnote{$\dagger$}   \\
%col1 text & col2 text  & col3 text  & col4 text  & col5 text   \\
%col1 text & col2 text  & col3 text  & col4 text  & col5 text\tnote{$\ddagger$}   \\
\bottomrule
\end{tabular*}
\begin{tablenotes}
\item[$\dagger$] 2-7 keV 
\end{tablenotes}
\end{table}
\end{center}

The key performance requirements for the WFI driven by this ambitious science program are shown in Table~\ref{tab:wfi}. These requirements are enabled by a Large Detector Array (LDA) consisting of four large-format ($512\times512$ pixels) DEPFET devices, delivering a field of view of $40\times40$ arc min. The pixel size corresponds to $2.2$ arcsec on the sky, oversampling the expected angular resolution of the \textit{Athena} mirror to provide accurate imaging and source positioning. A fifth DEPFET comprises the Fast Detector (FD) that is optimized for bright sources and is typically used out of focus to improve count rate capability. The DEPFET devices are controlled and read out by custom ASICs (Application Specific Integrated Circuits). These feed their signals into a total of five Detector Electronics (DE) units that perform the on-board frame processing. The output from the DE boxes is combined and fed to the spacecraft by the Instrument Control and Power-distribution Unit (ICPU). 

The WFI is being developed by a consortium of Universities and scientific Institutes, from 10 ESA Member State and the USA and is led by MPE. The instrument is currently in Phase B (preliminary definition), having passed successfully through its PRR in December 2018. A conceptual design instrument has been developed demonstrating the basic feasibility of the system. In parallel three key technology developments have been performed. The performance of the detector system (DEPFET+ASIC) has been demonstrated, as has the real-time performance of the DE. The large area, thin optical blocking filter has also recently passed acoustic noise tests; an important milestone to achieve the required technology readiness level at adoption. 

\section{The X-ray Integral Field Unit}
\label{xifu}

\begin{figure}[t]
\centerline{\includegraphics[scale=0.6]{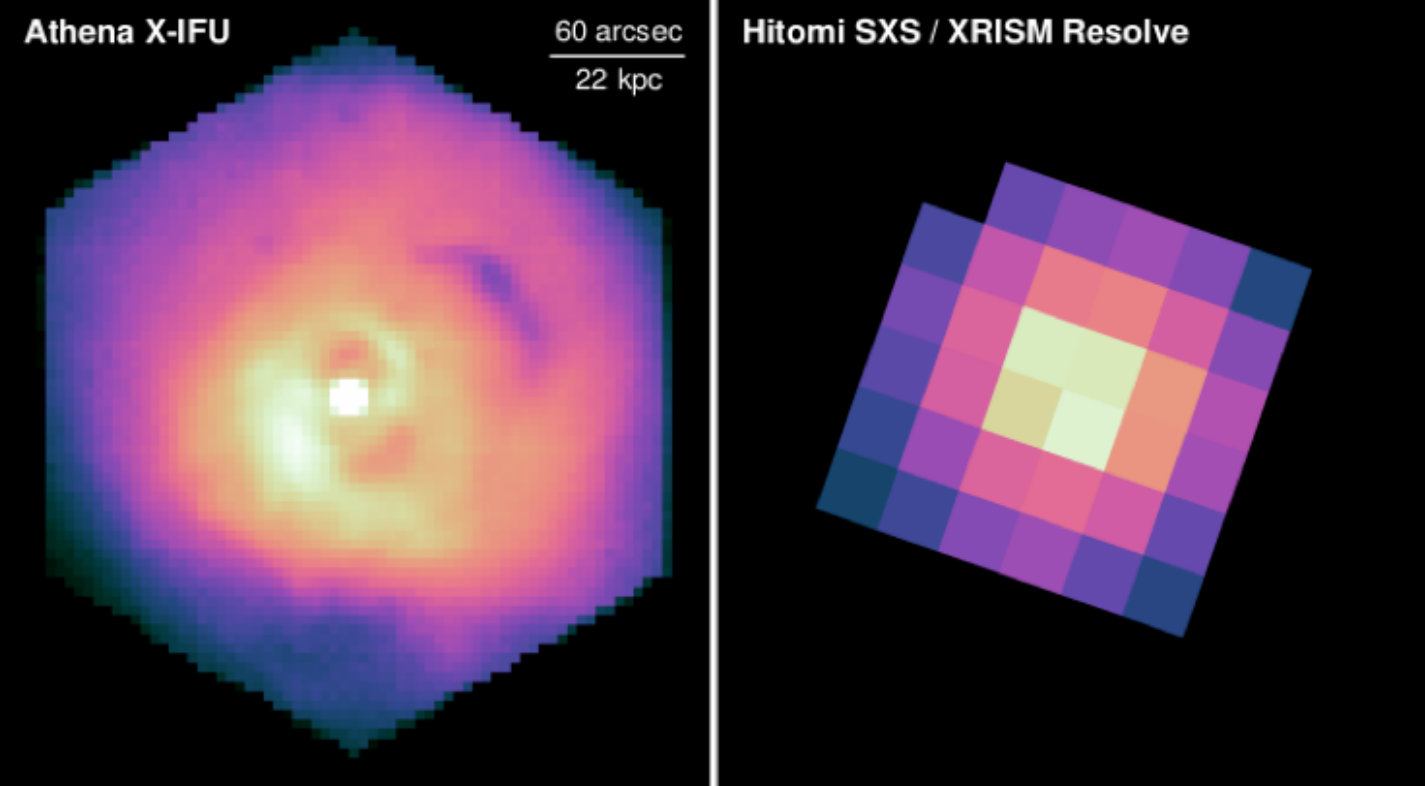}}
\caption{A simulated image of the Perseus cluster with X-IFU (left). The image of the same sky region with the Soft X-ray Spectrometer on board {\it Hitomi} on the same spatial scale as the \textit{Athena} image (right) \citep{hitomi_2016_nature}. Credit: Jeremy Sanders (MPE). Image composition: \textit{Athena} Community Office.}
\label{perseus_xifu_resolve}
\end{figure}

\begin{figure}[t]
\centerline{\includegraphics[scale=0.23]{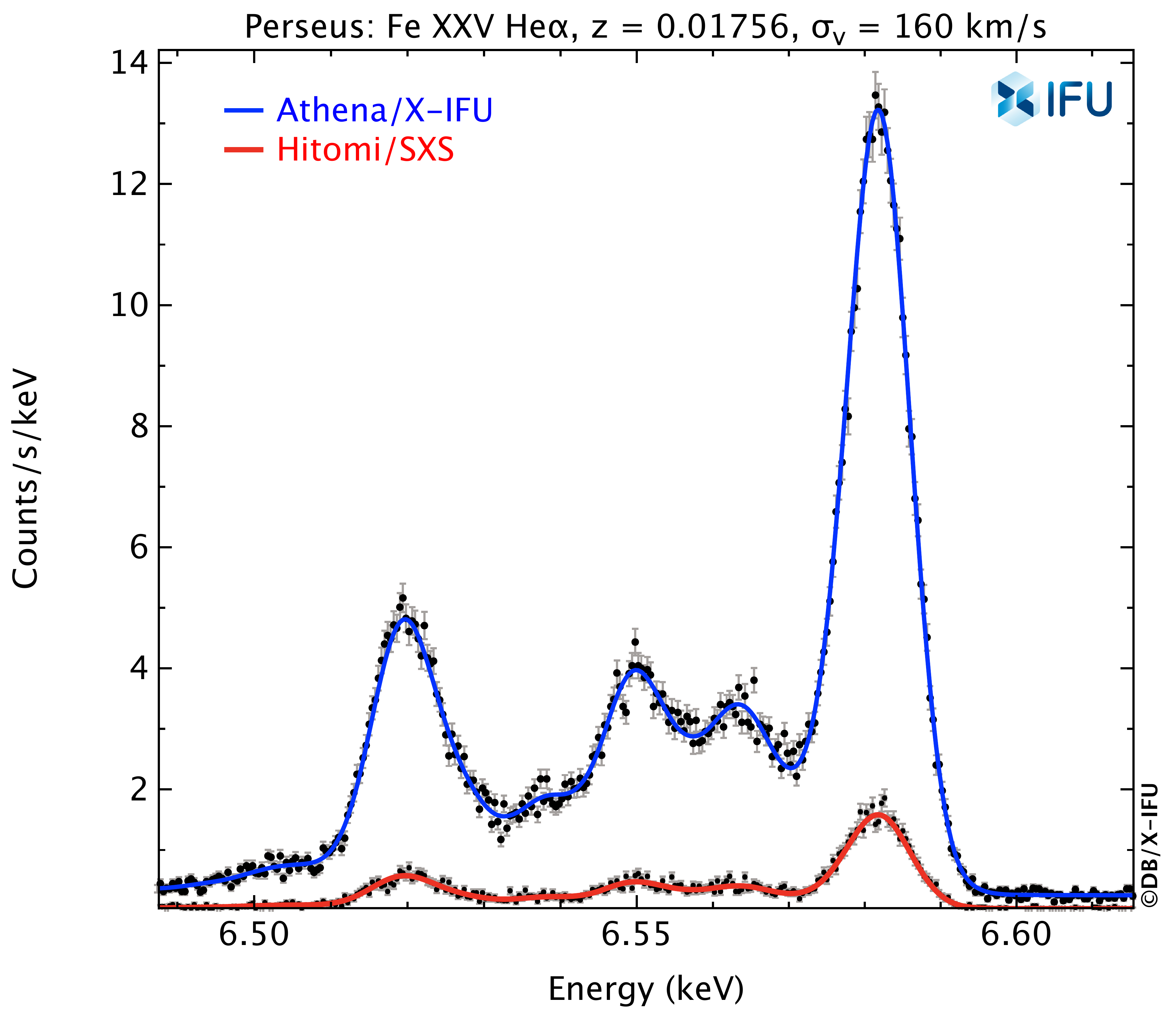}}
\caption{The \textit{Athena} X-IFU spectrum of the Fe XXV line compared to the one measured with the Hitomi Soft X-ray Spectrometer (SXS) (the  \textit{XRISM} \textit{Resolve} instrument will have the same performance as SXS). This illustrates the improvement in sensitivity due to the effective area increase between the two instruments. The input model is the one of a velocity- and thermally-broadened emission from collisionally-ionized diffuse gas (bapec in XSPEC terminology) with a plasma temperature of 4 keV temperature, metal abundance set to 0.7 and a Gaussian velocity broadening of 160 km/s.}
\label{spectrum_perseus_xifu_resolve}
\end{figure}

The \textit{Athena} X-ray Integral Field Unit is the cryogenic X-ray spectrometer of \textit{Athena} \citep{barret_spie_2016,barret_2018_spie,pajot_jltd_2018}. X-IFU will deliver spatially resolved X-ray spectroscopy, with a spectral resolution requirement of 2.5 eV up to 7 keV over an hexagonal field of view of 5 arc minutes (equivalent diameter). Key performance parameters relevant to the study of hot astrophysical plasma in extended sources (ie ignoring point source requirements) are summarized in Table \ref{tab1_xifu}. The leap of capabilities provided by \textit{Athena} with respect to  \textit{XRISM} is illustrated in Figures \ref{perseus_xifu_resolve}\ and \ref{spectrum_perseus_xifu_resolve} where the X-IFU image and spectrum of the Perseus cluster core are compared with those previously recorded with the Hitomi Soft X-ray Spectrometer \citep{hitomi_2016_nature}. 
\begin{center}
\begin{table}[t]%
\centering
\caption{X-IFU key performance requirements for studying hot astrophysical plasma.\label{tab1_xifu}}%
\tabcolsep=0pt%
\begin{tabular*}{20pc}{@{\extracolsep\fill}l@{\extracolsep\fill}}
\toprule
\textbf{Performance parameters}  \\
\midrule
Energy range : 0.2--12.0 keV \\
Spectral resolution : 2.5 eV (up to 7.0 keV) \\
Field of view : 5 arc minutes (equivalent diameter) \\
Pixel size : $\sim 5$ arc seconds \\
Non X-ray background : $5 \times 10^{-3} \rm ~counts/s/cm^2/keV$\tnote{$\dagger$}   \\
2.5 eV throughput (broadband, extended source) : 80\%\tnote{$\ddagger$} \\
%
%col1 text & col2 text  & col3 text  & col4 text  & col5 text\tnote{$\dagger$}   \\
%col1 text & col2 text  & col3 text  & col4 text  & col5 text   \\
%col1 text & col2 text  & col3 text  & col4 text  & col5 text\tnote{$\ddagger$}   \\
\bottomrule
\end{tabular*}
\begin{tablenotes}
\item[$\dagger$] In the 2-10 keV band.
\item[$\ddagger$] for a flux of $2\times10^{-11}$ ergs/s/cm$^2$/arcmin$^2$ (0.2-12 keV).
\end{tablenotes}
\end{table}
\end{center}

The prime detector is made of a large format array of Molybdenum Gold transition edge sensors coupled to absorbers made of Au and Bi to provide the required stopping power \citep{miniussi_jltd_2018}. The pixel size corresponds to slightly less than 5 arc seconds on the sky, thus matching the angular resolution of the X-ray optics. A large part of the X-IFU related \textit{Athena} science objectives relies on the observation of faint extended sources (e.g. hot gas in cluster of galaxies to measure bulk motions and turbulence or its chemical composition), imposing the lowest possible instrumental background. This is achieved by the addition of a second cryogenic detector underneath the prime focal plane array. This way non-X-ray events such as particles can be vetoed using the temporal coincidence of detecting energy in both detectors simultaneously. The focal plane array, the sensors and the cold front end electronics are cooled at a stable temperature less than 100 mK by a multi-stage cryogenic chain, assembled by a series of mechanical coolers, with interface temperatures at 15 K, 4K and 2K and 300 mK, pre-cooling a sub Kelvin cooler made of a $^3$He sorption cooler coupled with an Adiabatic Demagnetization Refrigerator. The multiplexed readout of the sensors is achieved in the frequency domain, while the processing of each event is performed on-board, through precise reconstruction of the pulses. Calibration data are acquired along each observation from modulated X-ray sources to enable the energy calibration required to reach the targeted spectral resolution. Although an integral field unit where each and every pixel delivers a high resolution X-ray spectrum, the defocussing capability of the \textit{Athena} mirror will enable the focal beam to be spread over hundreds of sensors. The X-IFU will thus be able to observe very bright X-ray sources \citep{peille_jltd_2018}. It will do so either with the nominal resolution, e.g. for detecting the baryons thought to reside in the Warm Hot Intergalactic Medium, using bright gamma-ray burst afterglows, as background sources shining through the cosmic web, or with a spectral resolution of 3-10 eV \citep[for a detailed explanation of the on-board pulse reconstruction for X-IFU]{peille_jltd_2018}, e.g. for measuring the spins and characterizing the winds and outflows of bright X-ray binaries at energies where their spectral signatures are the strongest (above 5 keV). 

The X-IFU will be provided by an international consortium led by France (IRAP and CNES), the Netherlands and Italy, with further ESA member state contributions from Belgium, Czech Republic, Finland, Germany, Ireland, Poland, Spain, Switzerland and key contributions from Japan and the United States. X-IFU proceeds now into Phase B after successfully passing its PRR. Priorities for the Phase B activities include any further optimizations of the cryogenic chain, the  readout chain (e.g., the multiplexing factor), a better assessment of the perturbations induced by micro-vibrations and electromagnetic interferences, as well as the consolidation of the interfaces with the science instrument module. These activities are supported by a vigorous technology development plan, required to bring all critical X-IFU technologies to a sufficient level of readiness for the adoption of \textit{Athena}. The development plan includes building a cooling chain providing a stable $\sim 100$ mK interface, for the testing of a functional demonstrator of the focal plane assembly. In parallel, the development of the multiplexed readout electronics should demonstrate its ability to meet the 2.5 eV resolution requirement after a careful optimization of the sensor functional parameters, e.g. pixel speed. The higher level of definition of the instrument and its main sub-system enables also the last round of performance optimization to be performed, most notably towards improving the instrument quantum efficiency at both 1 keV and 6-7 keV, which are two driving energies for the X-IFU science, e.g. for the Warm-Hot Intergalactic Medium (WHIM) detection and for measuring bulk motions in clusters. 

\section{Breakthroughs expected from \textit{Athena} on the astrophysics of extended X-ray sources}\label{sec5}
Having introduced the \textit{Athena} mission and its payload, let us now review the breakthroughs expected from \textit{Athena} on the astrophysics of extragalactic extended X-ray sources. 

\subsection{Hot extragalactic plasma}
\label{diffextragal}
The extragalactic sky contains many diffuse sources in a variety of shapes and sizes. Recent progresses on the study of diffuse emission from galaxies have been presented during the conference by \cite{li_tp,jimenez_gallardo_tp,paggi_tp,Vijayan_tp,Wezgowiec_tp,Zhang_tp,Zhang_tp_1}, and specifically, on the study of galaxy clusters by \cite{mernier_tp,pinto_tp,ettori_tp,Gaspari_tp,Hu_tp,Lagana_tp,Mao_tp,Caglar_tp,Ota_tp}. For recent reviews, see \cite{walker_2019_ssr,mernier_2018_ssr,pratt_2019_ssr,simionescu_2019_ssr}.

\begin{figure*}[!t]
\centerline{\includegraphics[scale=0.65]{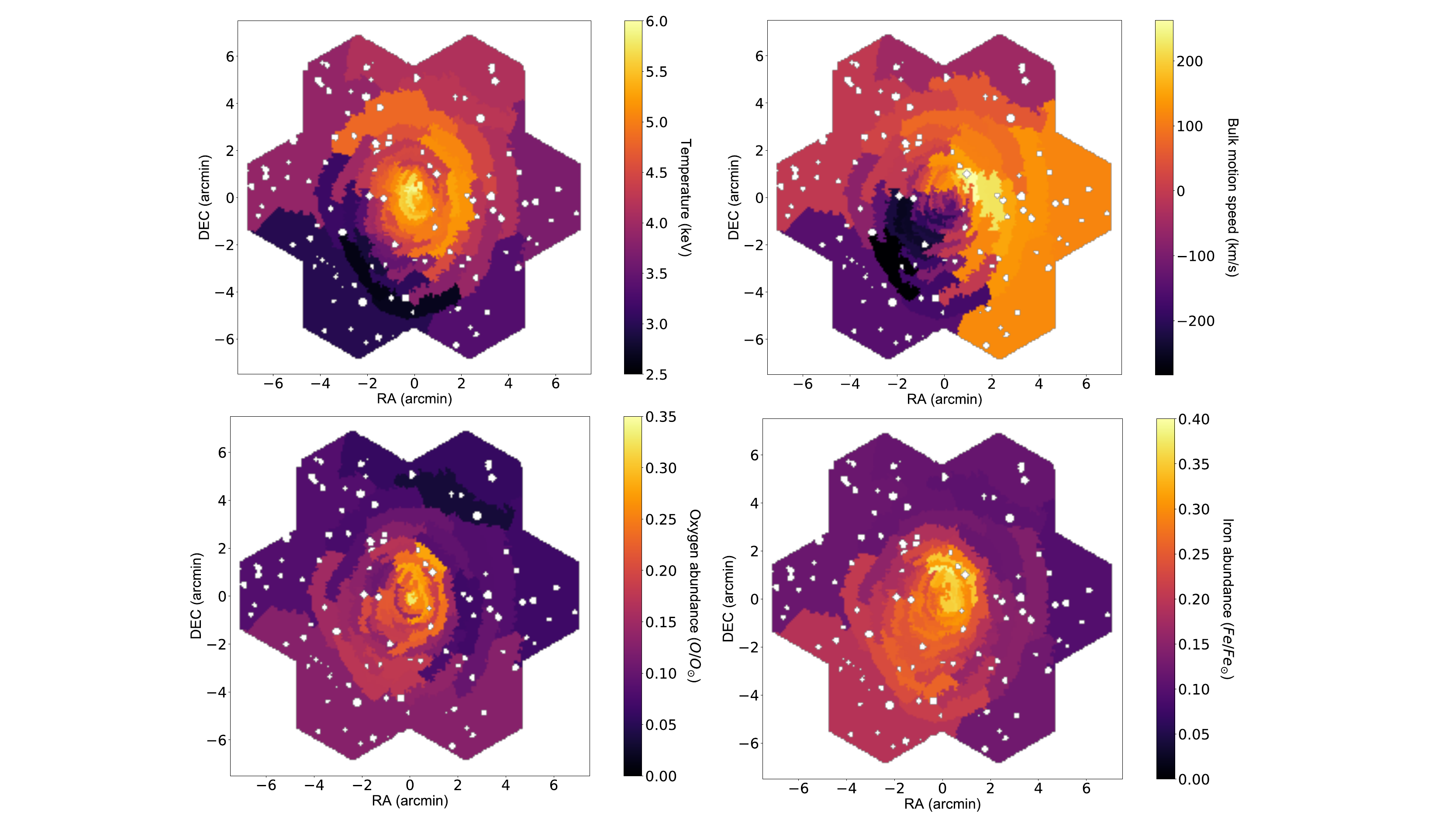} }
\caption{Physical parameter maps reconstructed from seven 100ks X-IFU observations of a galaxy cluster over regions of S/N~300 (~90,000 counts), simulated using the instrument end-to-end simulator SIXTE \citep{dauser_2019_aa}. The cluster is located at z=0.1, with $\rm R_{500} = 1.1$ Mpc and $\rm T_{500} = 4.2$ keV. Input clusters are taken from large-scale hydrodynamical simulations performed using the GADGET-3 smoothed-particle hydrodynamics code. (From top left to bottom right) Spectral-like temperature (keV), bulk velocity deduced from line-shift with respect to the cluster's average redshift (km/s), emission-measure-weighted oxygen and iron abundance (with respect to solar). The image is taken from \cite{cucchetti_2018_aa}, and references therein. White holes in the maps are associated with subtracted point sources.}
\label{fig_cucchetti}
\end{figure*}

The majority of extragalactic diffuse X-ray sources seen so far are thermal: hot gas trapped in the dark matter gravitational potential wells of clusters and massive galaxies. Some, however, are nonthermal and produced by relativistic particles jetted into space by Active Galactic Nuclei (AGN). Clusters of galaxies are the largest self-gravitating structures in the Universe. The mass of baryons they contain is by far dominated by the hot intracluster gas which is directly observable by its X-ray emission. Clusters are continuing to form at the present epoch. The mass function of clusters (number of clusters per unit mass) and its growth with redshift is sensitive to cosmology. When \textit{Athena} flies, the cosmological parameters describing the evolution of the Universe as a whole will be tightly constrained, e.g., through the eROSITA and Euclid missions. However, understanding the physics governing the assembly of the baryons across cosmic time will remain to be addressed by \textit{Athena} \citep{ettori_sp_2013,pointecouteau_sp_2013}. 

As shown in Figure \ref{fig_cucchetti}, \textit{Athena} with its unprecedented collecting area and combination of high spectral and angular resolution will enable to study how the baryonic gas accretes and evolves in the dark matter potential wells by measuring motions and turbulence in the intra-cluster medium \citep{cucchetti_2019_aa,clerc_2019_aa,rocarelli_2018_aa,bulbul_2019_usds}. \textit{Athena} will do so by resolving the accreting region both spatially and spectroscopically, probing the physical state of the X-ray emitting plasma \citep[and references therein]{ettori_sp_2013}. In addition as part of the survey to be conduced with WFI, \textit{Athena} should be able to determine how and when did the first galaxy groups in the Universe, massive enough to bind $10^7$ K gas, form \citep{pointecouteau_sp_2013}.

The temperature, density and metallicity distribution are a fossil record of the growth and past interactions of the cluster and its member galaxies \citep{ettori_sp_2013}. For the metalicity, recent work from Hitomi and XMM-Newton data have shown that the solar composition of the intra-cluster medium of Perseus is a common feature in nearby cool-core systems \citep{hitomi_nature_2017, mernier_2018_mnras}. The enrichment of the intracluster gas is also likely to have occurred at early epochs, as the protocluster formed. 

%The levels of bulk motions in the gas are linked to sloshing which is produced by minor mergers with subclusters. The cold fronts produced by this motion of the intracluster gas can be followed out to the virial radius, which means they last billions of years. Turbulence in the intracluster gas is linked to more major mergers, as well as energy feedback in which jetted AGN, particularly in the central galaxy, stir the central gas up (see Gaspari, this volume).

%\begin{figure}
%\centerline{\includegraphics[scale=0.4]{Perseus_v6.pdf}}
%\caption{Chandra gradient-filtered X-ray image of the  cool core of the Perseus cluster (Sanders) overlaid with Hitomi spectrum covering Fe XXV He{$\alpha$} and Fe XXVI Ly{$\alpha$}.}
%\end{figure}

\begin{figure}
\centerline{\includegraphics[scale=0.1]{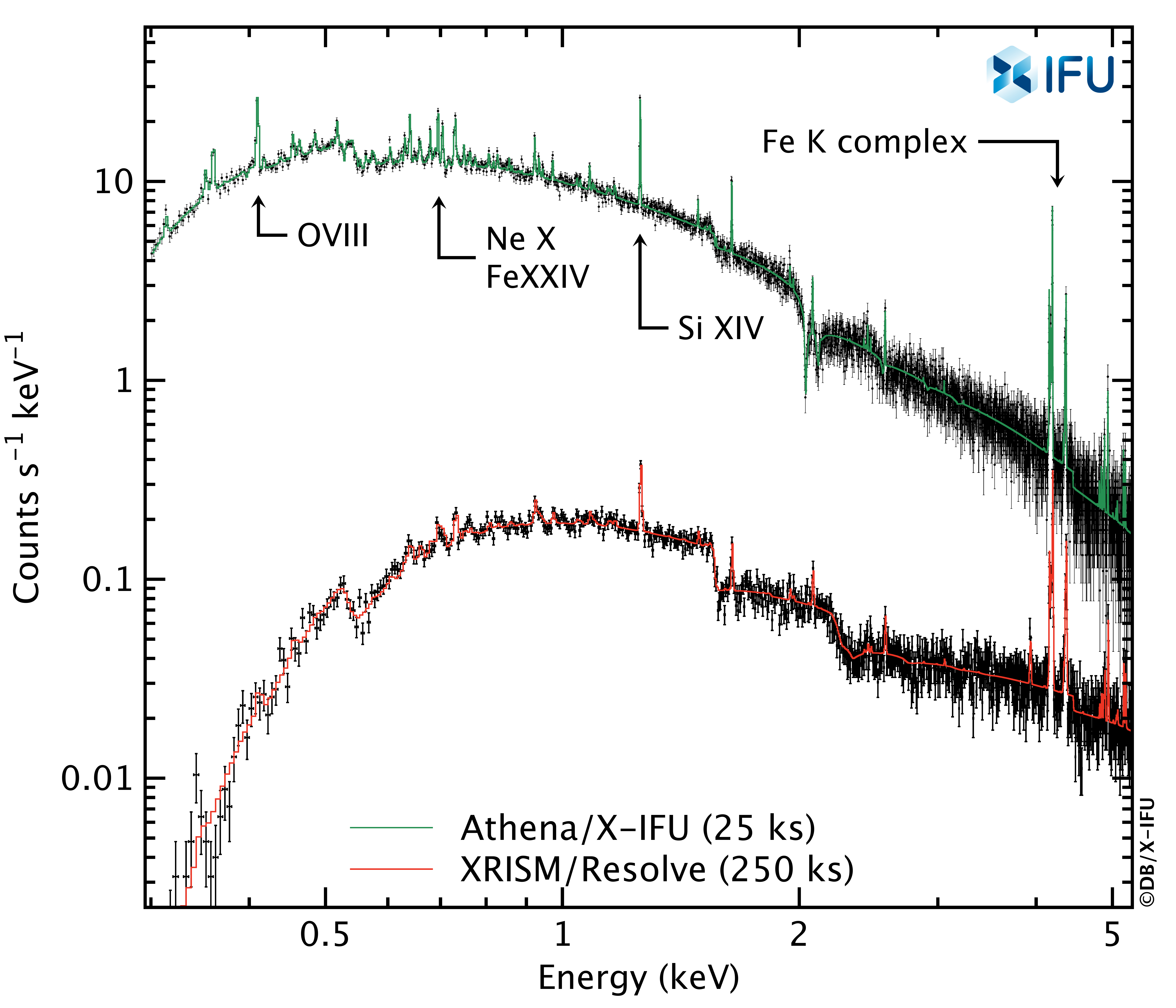} }
\caption{Simulations of the X-ray spectrum of the Phoenix cluster ($z=0.59$); based on XMM-Newton RGS \citep{pinto_2018_mnras} for the X-IFU and \textit{Resolve} spectrometers. The integration time is 25 ks and 250 ks, respectively. The  \textit{XRISM} \textit{Resolve} response has been assumed to be the same as the one of the Hitomi/SXS instrument (with the gate valve open). Some indicative emission lines are reported. Data courtesy of C. Pinto. }
\label{fig_phenix}
\end{figure}

The major leap in capabilities provided by \textit{Athena}, even over  \textit{XRISM} (see Figure \ref{fig_phenix})  will enable to access the properties of the hot gaseous atmosphere of groups and clusters out to their formation epoch, when both star formation and black hole activity were at their maximum. This will reveal how the bulk of heavy elements was generated and distributed in the hot intergalactic medium, and how important AGN and supernova feedback was at high redshift when groups and clusters were born (see Mernier and Ettori this volume and see also the recent work of \cite{cucchetti_2018_aa}). From observations of low-z systems (with WFI and X-IFU) associated with studies of $z=1-2$ group and cluster properties, \textit{Athena} will bridge the gap between the era of galaxy formation and early evolution (i.e., beyond $z\sim 2-3$; traced by the Ly-$\alpha$ forest) and the lower redshift Universe where large-scale structures are assembling and virialising \citep[and references therein]{pointecouteau_sp_2013}.

AGN feedback provides sufficient power to offset radiative cooling of the central gas in cool-core clusters, and is a key ingredient in current models of galaxy formation and cluster evolution. If AGN feedback were not present in cool-core clusters the intracluster medium would rapidly cool and make central cluster galaxies have much higher stellar masses than are seen. Some cooling and star formation are seen but the rates are much lower than would be expected without feedback. The low level of turbulence measured in the Perseus cluster by Hitomi \citep{hitomi_2016_nature} and in some other clusters with the XMM-Newton RGS (see Pinto, this volume) leaves a major energy flow problem: how does the energy flow from the AGN throughout the cluster core? The turbulence is too low to be effective by itself, so is the flow due to sound waves, gravity waves, cosmic rays, or something else? The \textit{Athena} X-IFU, thanks to its unprecedented combination of spatial spectral resolution and effective area, will enable a robust physical understanding of this feedback mode, through the first spatially-resolved kinematic measurements on relevant spatial scales of the hot gas in galaxy, group and cluster haloes as it absorbs the impact of AGN jets, and via the mapping of thermodynamic conditions on scales well-matched to the jets, lobes and gas disturbances produced by them. It will therefore be possible to precisely quantify how jet energy is dissipated and distributed in group and cluster gas, and how the feedback loop operates in group/cluster cores to regulate gas cooling and AGN fuelling. Similarly, \textit{Athena} will also establish firmly the cumulative impact of powerful radio galaxies on the evolution of baryons from the epoch of group/cluster formation to the present day \citep[and references therein]{croston_sp_2013}.

%The intracluster gas is magnetized, as revealed by Faraday Rotation studies as well as the presence of radio haloes. This greatly complicates our understanding of the microphysics and transport phenomena such as viscosity and conductivity in the gas.  

It is worth stressing that what is seen now in clusters is only a broad-brush guide. The nearest and brightest clusters, e.g. Perseus, Virgo and Centaurus, show rich behaviour in temperature and density structure as well as metallicity. A first glimpse to this complex phenomenology will be provided by  \textit{XRISM}, whose spatial resolution is, however, limited by the angular resolution of the telescope (~1') and the pixel size of the {\it Resolve} spectrometer (~30"). \textit{Athena}, with its deeper and sharper view  will open up exploration of the present and past behaviour of these giant components of the Universe. 

\begin{figure}[!h]
\centerline{\includegraphics[scale=0.225]{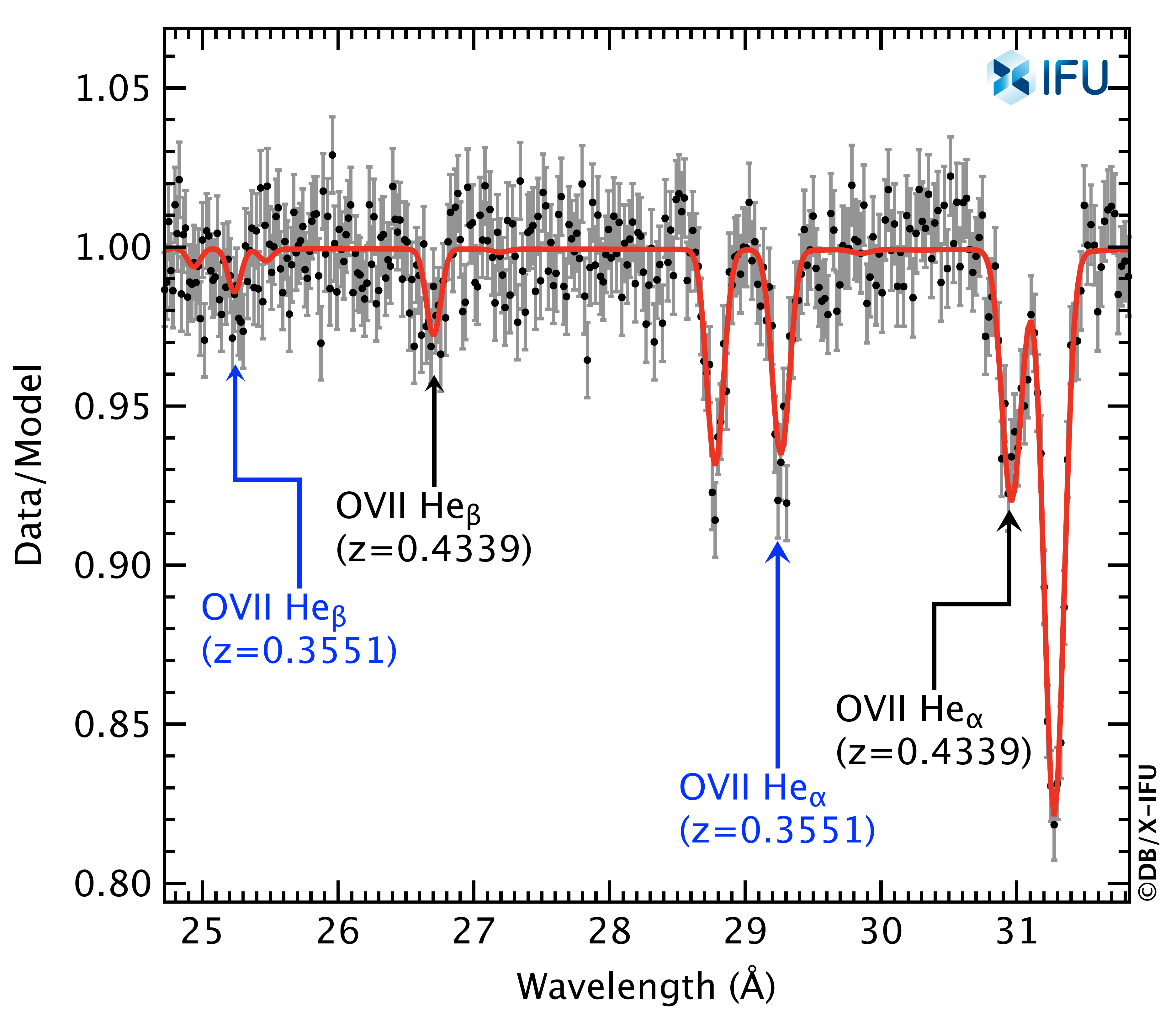} }
\caption{Simulated X-IFU spectrum of two intervening absorbers at z = 0.3551 and z=0.4339. The plot shows the ratio of the X-IFU data of the blazar 1ES 1553+113 with the local best-fitting continuum model, to highlight the two OVII He$\alpha$  and He$\beta$  absorption lines. The simulation made use of the best fitting parameters derived from the long XMM-Newton RGS observations reported in \cite{nicastro_nat_2018}.}
\label{fig_nicastro}
\end{figure}

%All massive galaxies have, to varying degrees, hot gaseous haloes. They too have similar, scaled down, properties to clusters and groups.  Understanding and accounting for this gas accurately is important for understanding galaxy evolution (e.g. Li, this volume). 

Complementary to the study of the intergalactic plasma permeating galaxy
groups and clusters, \textit{Athena} will unveil the location, chemical composition,
physical state and dynamics of the thermodynamically dominating phase of
baryons in the local Universe (z$<$1), residing in unvirialized, filamentary
structures, the so-called WHIM. According to
large-scale cosmological simulations, these baryons are supposed to constitute
a dilute gaseous phase at temperatures $\ge 10^{5.5}$~K. Sensitive,
high-resolution measurements in soft X-rays are the only tool to experimentally
detect this component. The WHIM has been historically rather elusive. Almost one
half the the baryon census in the local Universe seemed to be missing
see,  e.g. \cite{nicastro_an_2017}. The first solid proof of its existence
was gathered by an ultra-deep ($\sim$1.5~Ms) dedicated observational campaign
with the XMM-Newton RGS \citep{nicastro_nat_2018}. The unprecedented combination
of effective area and energy resolution will allow the \textit{Athena} X-IFU to address
the cosmic distribution of the WHIM; derive its physical state and composition;
investigate the ultimate fate of the missing baryons in galactic halos,
including those coming from our own Galaxy; and to estimate how much material
is accreted, blown out, and locked-up temporarily in stars, providing an
experimental estimate of the relative contribution of accretion versus
outflows in structure formation \citep[and references therein, see Fig. \ref{fig_nicastro}]{kaastra_2013_sp}.

\subsection{Hot galactic plasma}
\label{diffgal}
Hot diffuse plasma can be observed in many parts in our Galaxy, from supernova remnants to the galactic central lobe and galactic plane. Diagnosing their spatially resolved spectra is essential to reveal the composition, properties  and origin of the hot plasma. Their studies in X-rays convey prime information on the energy output from the supernovae, stellar wind of massive stars, properties of the interstellar medium, or on the activity of the Galactic center massive black hole Sgr A* \citep[and references therein]{decourchelle_sp_2013}. Recent progresses in the field in terms of observations and modelling have been presented during this conference by \cite{ferrand_tp,yamaguchi_tp,an_tp,matsumara_tp,zhou_tp,greco_tp}.

\begin{figure}[t]
\centerline{\includegraphics[clip,width=0.48\textwidth]{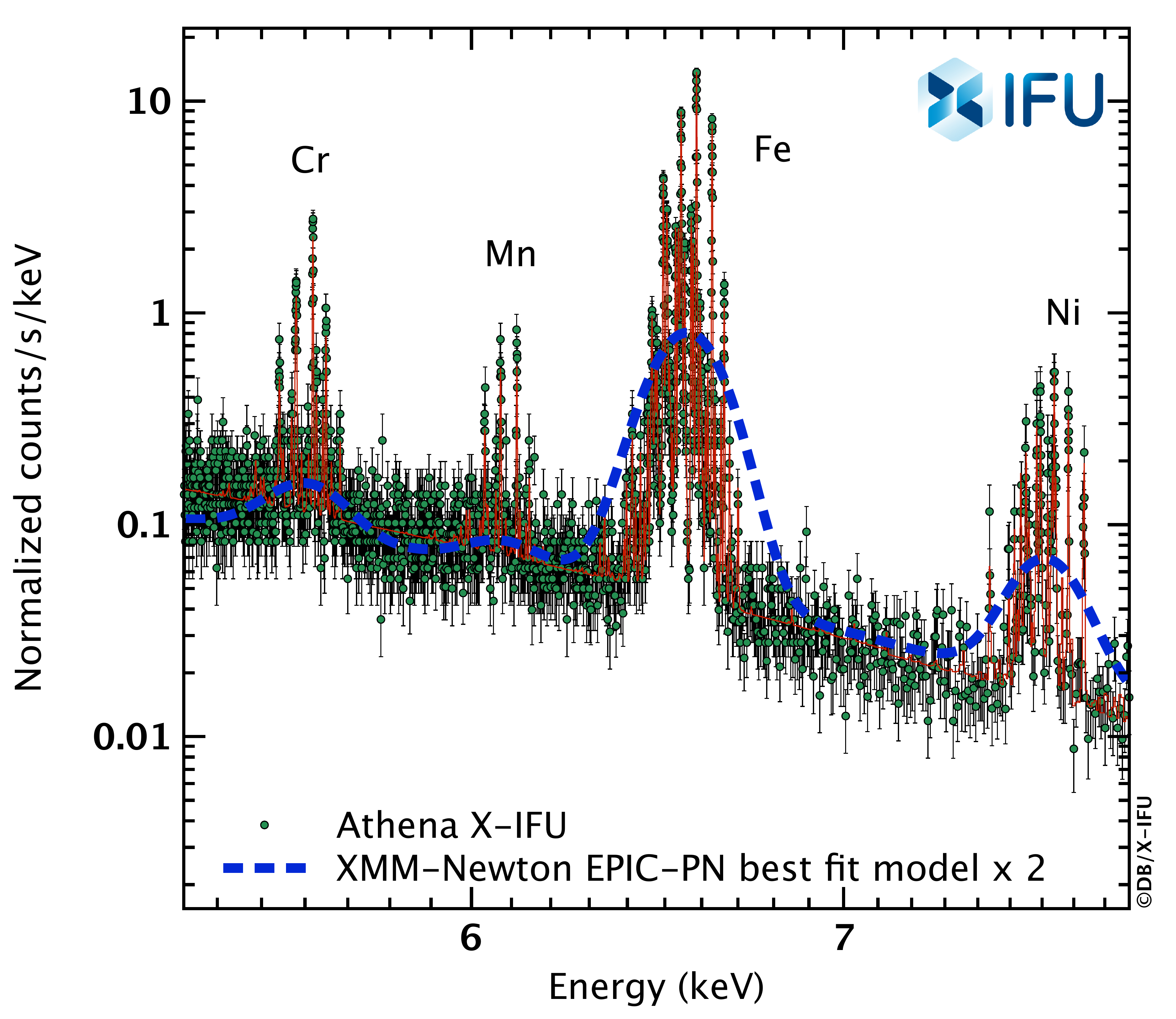}}
\caption{The \textit{Athena} X-IFU spectrum of the Fe group elements in 3C397, a few thousand year old galactic supernovae remnant of a type Ia supernovae. The XMM-EPIC best fit spectral model (scaled up by a factor of 2) is shown to illustrate the power of high resolution spectroscopy. This figure is adapted from \cite{williams_2019_arxiv}. Such a measurement would enable direct comparisons with hydrodynamic modeling of supernova explosion, as required to bridge the gap between the stellar progenitor, the explosion and the remnant \citep{williams_2019_arxiv}.}
\label{fig_snr}
\end{figure}

This energy and heavy element injection has a considerable impact on  the interstellar medium and the overall dynamics of the Galaxy, generating heating, turbulence, heavy element enrichment, particle acceleration and magnetic field amplification at high Mach number shocks. Both thermal and nonthermal X-ray emission is generated and provides unique diagnostic on the plasma properties and the physical processes at work. Emission lines can be observed in X-rays from various elements synthesized in supernovae and their progenitors (oxygen, neon, magnesium, silicon, sulphur, argon, calcium, iron) giving access to the composition of the hot plasma as well as the element ionization states \citep{williams_2019_arxiv} (see Fig. \ref{fig_snr}). The characterization of the synchrotron emission from TeV electrons provides information about the properties of the particle acceleration process. X-ray absorption spectroscopy allows to study not only the interstellar gas, but gives also access to the interstellar dust composition and properties e.g., \cite{2016Schulz_ApJ}.

With its unrivaled performances, \textit{Athena} will provide breakthroughs in this field on a number of galactic sources and diffuse components. 

For supernova remnant, spatially-resolved high spectral spectroscopy provided by \textit{Athena} X-IFU will enable the derivation of 3D maps for the different synthesized elements in the ejected material. The morphology of the supernova remnants is indeed a clue to SN type, explosion mechanism and progenitor; e.g. see review by \cite{Lopez_Fesen_2018_SSRv}. More precisely, the morphology and asymmetries in the ejection of the supernova material are intimately related to the properties of the explosion mechanism, whose understanding is still uncomplete. \textit{Athena} X-IFU will provide measurement of the velocities in the line of sight of the different elements produced by the supernova, as well as their properties including those of rare elements (Ca, Mn, Ti, Ni).

The characterization of the observed TeV electron synchrotron emission provides information about the properties of the particle acceleration process. While the WFI will enable mapping on large scale the properties of the synchrotron and thermal emissions at a spatial resolution of less than 10 arsec, X-IFU will enable specific measurement of the post-shock temperature in regions dominated by nonthermal emission, thanks to its high resolution spectroscopy. This will give constraints on the level of feedback of ion acceleration on the shock and hydrodynamic structure, e.g. Ferrand and Dwarkardas (this volume).

\begin{figure}[t]
\centerline{\includegraphics[scale=0.7]{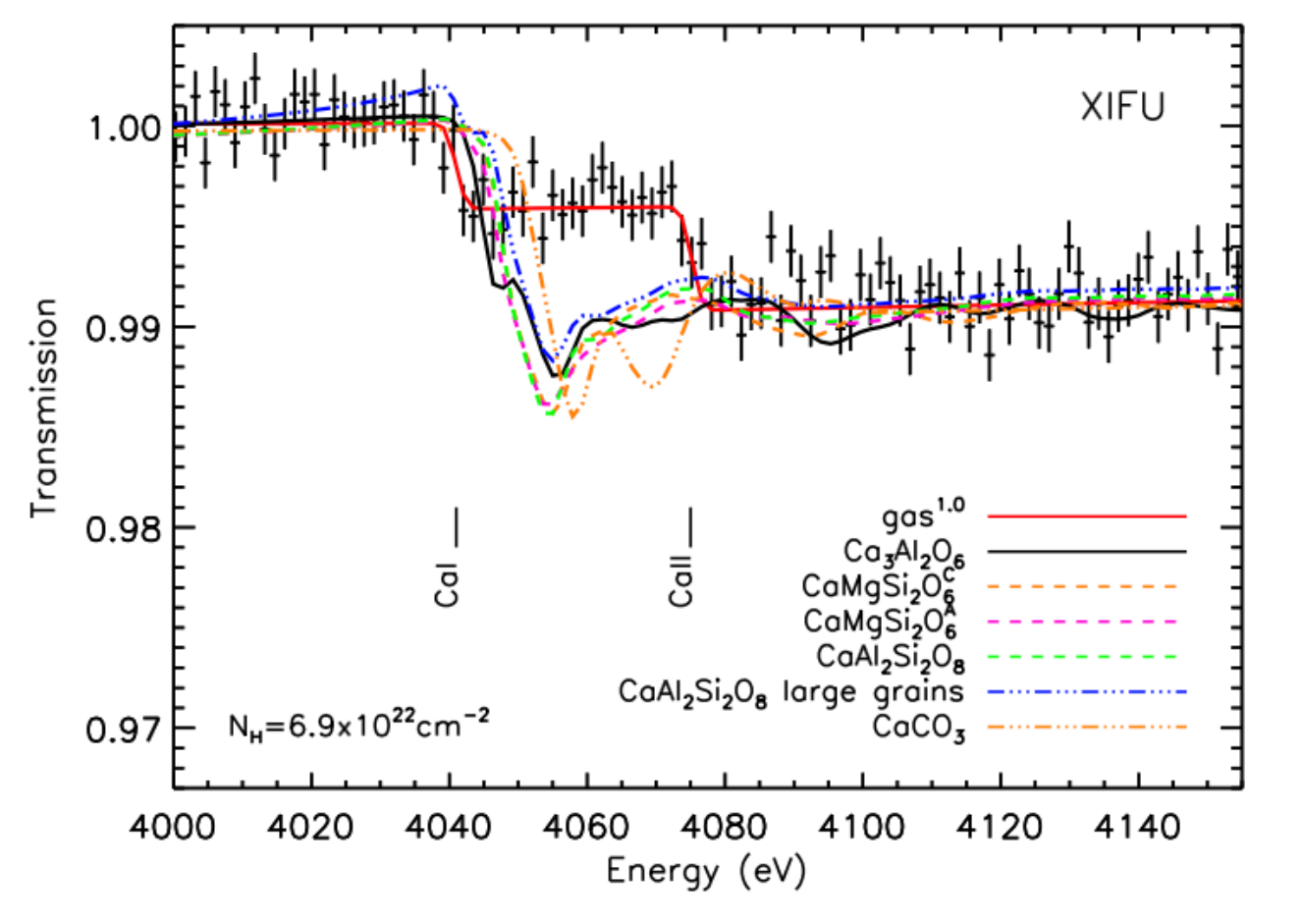}}
\caption{X-IFU simulation of the absorption feature in the line of sight of the bright XRB GX340+00 for an exposure time of 400ks: the calcium K-edge is shown for different calcium species with a dust depletion of 100\%. From \cite{Costantini_2019_aa}.}
\label{fig_dust}
\end{figure}

The detailed characterization of the plasma properties and notably in terms of ionizing or recombining plasma, with the sensitivity, spectral and spatial resolution of \textit{Athena}, will provide specific diagnostics of the environment related to the progenitor (stellar wind, shell) and the interaction with a cloudy interstellar medium (see Yamaguchi, this volume). Recombining plasmas in supernova remnants have been first revealed by enhanced Ly alpha lines in the early 2000 in IC443 with ASCA. Since then, a number of supernova remnants have been identified with overionized plasmas, generally in mixed morphology remnants, whose origin is still not resolved. \textit{Athena} will provide unique information having the sensitivity, a high spectral resolution and a spatial resolution enabling measurement in regions of  typically tens of arc seconds.

The study of supernova remnants in nearby galaxies (see Sasaki, this volume) will strongly benefit of \textit{Athena} allowing a fine characterization of the spectra and properties of the population in terms of composition, supernova type and progenitors.
Regarding pulsar wind nebulae, \textit{Athena} will provide a major leap allowing a deep investigation with X-IFU of any weak thermal emission, whose origin relates to the progenitor and environment of the supernova, and a better census of this population by probing the faint structures with the WFI. On Wolf-Rayet bubbles and planetary nebulae, only few objects are currently detected in X-rays and already open a number of issues on their temperature and composition. \textit{Athena} provides a bright future to study the hot plasma in these sources.

X-ray spectroscopy of the cold interstellar medium will probe dust size distribution, composition and geometry, an important component of the universe, present in every stage of the stellar evolution (see Pellegrini this volume, see also \cite{Costantini_2019_aa} and Fig. \ref{fig_dust}).

Another strong asset of \textit{Athena} will be the study of the hot interstellar medium at the Galactic centre \citep{cappi_sp_2013}, a complex ecosystem with a supermassive black hole Sgr A*, molecular clouds, supernova remnants, non-thermal filaments, and chimneys as recently shown by \cite{2019Ponti_Nature}. With the WFI, \textit{Athena} will map nearby remnants and extended diffuse emission of hot plasma to diagnose particle acceleration, plasma enrichment and the properties of the hot interstellar medium phase.

\section{Conclusions}
\textit{Athena} is our next large X-ray observatory. It has revolutionary X-ray capabilities for spatially resolved high-resolution spectroscopy and wide-field spectral imaging, and will provide breakthrough observations of hot astrophysical plasma. Among the challenges to be addressed by \textit{Athena}, one can cite calibration, background (and the contribution of charge exchanges, see Molendi and Betancourt-Martinez, this volume), atomic physics (see Foster, this volume and \cite{Betancourt_2019_baas}), and data analysis (see the presentation of Gastaldello, this conference) to cite a few. Those will require some efforts by the entire high-energy astronomy community. As far as \textit{Athena} is concerned, the successful end of phase A review for both instruments puts it on a safe path for a launch in the early 2030s. In order to support the ultimate adoption of \textit{Athena} in the ESA Science Program at the end of 2021, the \textit{Athena} Science Study Team will have to provide the so-called "Red Book", for which the science case of the \textit{Athena} will have to be updated. In the meantime, the two great XMM-Newton and \textit{Chandra} observatories will hopefully continue to produce excellent scientific results, while  \textit{XRISM} will open the window of non-dispersive high-resolution X-ray spectroscopy. For the most distant future, the youngest members of our community have already engaged in the voyage2050 exercise for the long term science planning of ESA to prepare the follow-up mission to \textit{Athena}, e.g. \cite{simionescu_voyage_2019} and \cite{uttley_voyage}.
\section{Acknowledgments}
The authors wish to thank the ESA study team, and the  \textit{Athena} working and topical panel members, who contributed to the elaboration of  the science case for  \textit{Athena}. We are also grateful to the Athena Community Office for supporting the  \textit{Athena} community. DB is grateful to Etienne Pointecouteau for comments on an earlier version of this paper and to C. Pinto, F. Nicastro and B. Williams for providing the data for figures \ref{fig_phenix}, \ref{fig_nicastro}, \ref{fig_snr}. DB, on behalf of the X-IFU consortium, is grateful to the CNES project team, led by Vincent Albouys, for managing the development of the instrument.

The authors thank an anonymous referee for a careful review of the paper. 
\bibliography{db}%
\jnlcitation{\cname{%
\author{Barret et al.},} (\cyear{2019}), 
\ctitle{The Athena space X-ray Observatory and the astrophysics of hot plasma, \cjournal{Astronomische Nachrichten / Astronomical Notes (AN) }, \cvol{TBD}.}}

\end{document}